\documentclass[sigconf]{acmart}

\AtBeginDocument{%
  \providecommand\BibTeX{{%
    \normalfont B\kern-0.5em{\scshape i\kern-0.25em b}\kern-0.8em\TeX}}}

\setcopyright{acmcopyright}
\copyrightyear{2021}
\acmYear{2021}

\acmConference[CIKM '21]{The 30th ACM International Conference on Information and Knowledge Management}{November 1--5, 2021}{Online}
\acmBooktitle{The 30th ACM International Conference on Information and Knowledge Management (CIKM '21), November 1--5, 2021, Online}
\acmPrice{15.00}
\usepackage{adjustbox}
\usepackage{CJKutf8}
\usepackage{dblfloatfix}
\usepackage{bm}
\usepackage{mathrsfs}

\copyrightyear{2021}
\acmYear{2021}
\setcopyright{acmcopyright}\acmConference[CIKM '21]{Proceedings of the 30th ACM International Conference on Information and Knowledge Management}{November 1--5, 2021}{Virtual Event, QLD, Australia}
\acmBooktitle{Proceedings of the 30th ACM International Conference on Information and Knowledge Management (CIKM '21), November 1--5, 2021, Virtual Event, QLD, Australia}
\acmPrice{15.00}
\acmDOI{10.1145/3459637.3481983}
\acmISBN{978-1-4503-8446-9/21/11}

\begin{document}
\begin{CJK*}{UTF8}{gkai}
\fancyhead{} 

\title{AliMe MKG: A Multi-modal Knowledge Graph for \protect\\ Live-streaming E-commerce}


% \author{Guohai Xu, Hehong Chen, Feng-Lin Li, Yunzhou Shi, Wei Zhou, Zhongzhou Zhao, Ji Zhang}
% \email{{guohai.xgh, hehong.chh, fenglin.lfl}@alibaba-inc.com}
% \affiliation{%
%   \institution{DAMO Academy, Alibaba Group}
% }

\author{Guohai Xu}
\authornote{Both authors contributed equally to this research.}
\email{guohai.xgh@alibaba-inc.com}
\author{Hehong Chen}
\authornotemark[1]
\email{hehong.chh@alibaba-inc.com}
\affiliation{%
  \institution{DAMO Academy, Alibaba Group}
}

% \author{Hehong Chen}
% \email{hehong.chh@alibaba-inc.com}
% \affiliation{%
%   \institution{DAMO Academy, Alibaba Group}
% }

\author{Feng-Lin Li}
\authornote{Feng-Lin Li is the corresponding author.}
\email{fenglin.lfl@alibaba-inc.com}
\author{Fu Sun}
\email{fusun.sf@alibaba-inc.com}
\affiliation{%
  \institution{DAMO Academy, Alibaba Group}
}

% \author{Fu Sun}
% \email{fusun.sf@alibaba-inc.com}
% \affiliation{%
%   \institution{DAMO Academy, Alibaba Group}
% }

\author{Yunzhou Shi}
\email{yunzhou.syz@alibaba-inc.com}
\author{Zhixiong Zeng}
\email{zengzhixiong.zzx@alibaba-inc.com}
\affiliation{%
  \institution{DAMO Academy, Alibaba Group}
}

% \author{Yunzhou Shi, Wei Zhou}
% \email{{yunzhou.syz,fayi.zw}@alibaba-inc.com}
% \affiliation{%
%   \institution{DAMO Academy, Alibaba Group}
% }

\author{Wei Zhou}
\email{fayi.zw@alibaba-inc.com}
\affiliation{%
  \institution{DAMO Academy, Alibaba Group}
}

\author{Zhongzhou Zhao}
\email{zhongzhou.zhaozz@alibaba-inc.com}
\affiliation{%
  \institution{DAMO Academy, Alibaba Group}
}

\author{Ji Zhang}
\email{zj122146@alibaba-inc.com}
\affiliation{%
  \institution{DAMO Academy, Alibaba Group}
}

\renewcommand{\shortauthors}{Xu, et al.}

\begin{abstract}
Live streaming is becoming an increasingly popular trend of sales in E-commerce. The core of live-streaming sales is to encourage customers to purchase in an online broadcasting room. To enable customers to better understand a product without jumping out, we propose \textbf{AliMe MKG}, a multi-modal knowledge graph that aims at providing a cognitive profile for products, through which customers are able to seek information about and understand a product. Based on the MKG, we build an online live assistant that highlights product search, product exhibition and question answering, allowing customers to skim over item list, view item details, and ask item-related questions. Our system has been launched online in the Taobao app, and currently serves hundreds of thousands of customers per day.
\end{abstract}

\ccsdesc[500]{Information systems~Graph-based database models}
\ccsdesc[500]{Computing methodologies~Information extraction}
\ccsdesc[500]{Computing methodologies~Matching}

% \keywords{datasets, neural networks, gaze detection, text tagging}
\keywords{Knowledge Graph, Multi-modality, Live-streaming, E-commerce}


\maketitle

\section{Introduction}

In recent years, live streaming is becoming an increasingly popular trend of sales in E-commerce. In a live streaming, an anchor will introduce the listed items~\footnote{We use items and products interchangeably in this paper.} in an attractive way, and offer certain discounts or coupons, to facilitate user interaction and volume of transaction. If a customer is interested in a particular item, s/he could further view its details, and ask specific questions. Being different from traditional purchasing process, where customers make purchasing decision through product search and evaluation on an E-commerce website, it is more natural for customers of live streaming to purchase within an online broadcasting room. 

To support this new purchasing mode, we need to provide rich and attractive information that details the important aspects of each product item in an online broadcasting room. It is better for such information to constitute a cognitive product profile, through which customers are able to truly understand a product: why they need, when and where can they use, how to use, what is the effect, etc. Also, as customers may have specific questions about an item during browsing, we also need to help an anchor to answer questions, to respond to users and provide a better user experience. 

To this end, we establish \textbf{AliMe MKG}, a multi-modal knowledge graph that centers on and aggregates rich information about items. Based on the knowledge graph, we further build an online live assistant that highlights product search, product exhibition and question answering, allowing customers to conveniently seek information in an online broadcasting room, including but not limited to skimming over item list, viewing item details, and asking item-related questions. 

Our system has been launched online in the Taobao APP, and currently serves hundreds of thousands of customers per day. In this paper, we introduce how we construct our multi-modal knowledge graph through cross-modal information fusion, and demonstrate its value through its application in the live assistant. 

% \section{System Overview}
% In this section, we first introduce the ontology of our knowledge graph, then describe the construction process, and at last show its applications for live streaming.

\section{Multi-Modal Knowledge Graph}
In this section, we introduce our multi-modal knowledge graph, including core ontology, construction process, experimental results and statistics.

\subsection{KG Ontology}

% \begin{figure}[t]
% 	\centering
% 	\includegraphics[width=\linewidth]{files/mining_process.pdf}
% 	\caption{The overview of AliMe MKG.}
% 	\label{fig:construction}
% \end{figure}

\begin{figure}[h]
	\centering
	\includegraphics[width=0.98\linewidth]{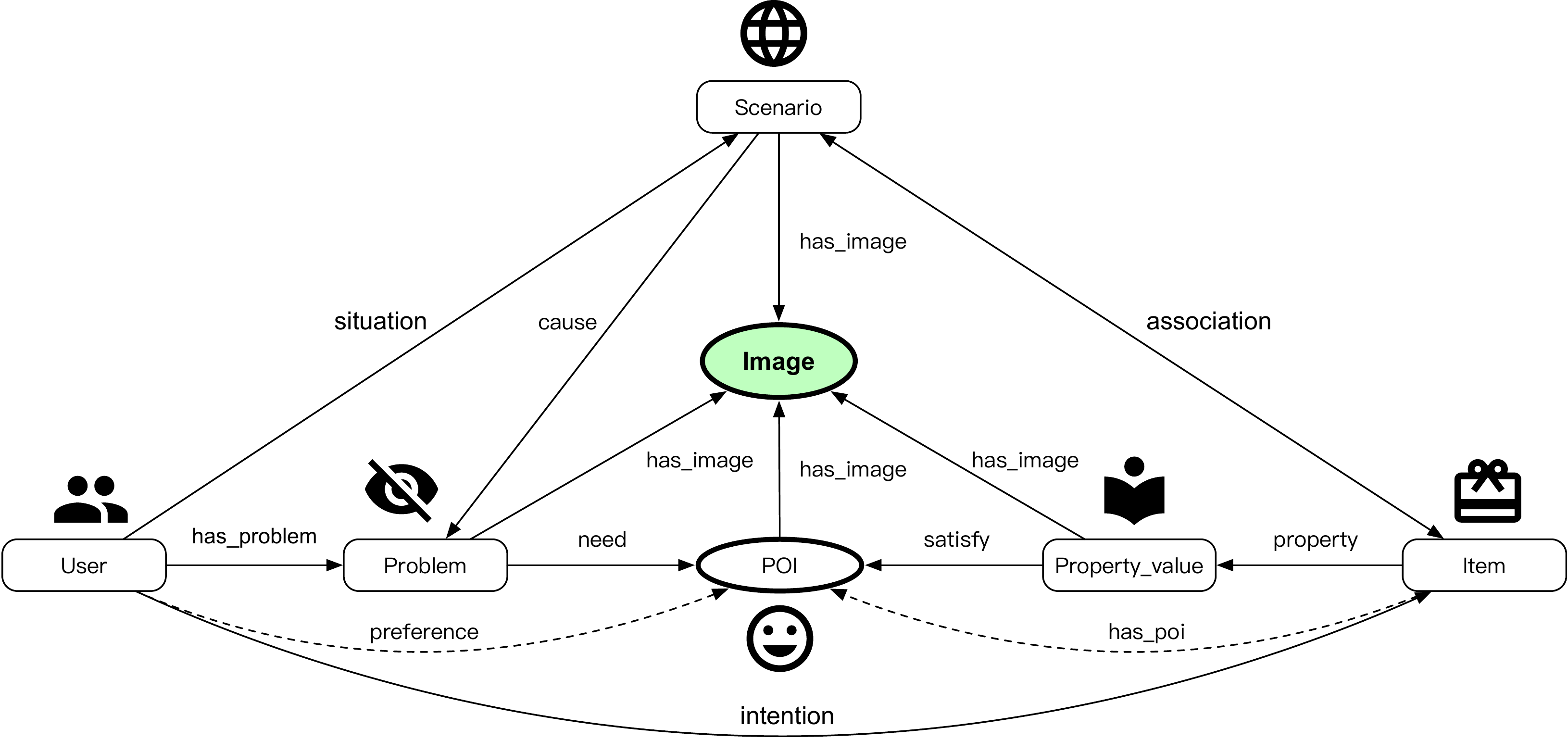}
	\caption{The core ontology of AliMe MKG.}
	\label{fig:kg_ontology}
\end{figure}

We show our core ontology in Figure~\ref{fig:kg_ontology}. Three commonly accepted concepts, namely ``User'', ``Item'' and ``Scenario'', are adopted from classic buying process: a user intent to buy some items at/for a certain scenario. The concept ``Property\_value'' captures property values of items (e.g., ``bisabolol 红没药醇'' is the property value of ingredient of ``cleansing foam 洁面泡沫''). ``Problem'' and ``POI (Point of Interest)'' are adopted from AliMe KG~\cite{DBLP:journals/corr/abs-2009-11684}, our previous domain knowledge graph: ``Problem'' refers to a problematic state that a user is at (e.g., ``pimple 长痘痘''); ``POI'' captures users’ need or solution to user problem (``antiacne 清痘抑痘''). The new concept ``Image'' is a visual entity that graphically represents a textual scenario, user problem, property value or POI, and is the key contribution of this work.

% The concept “IPV” captures property values of items(Item-Property-Value, e.g., “cleansing foam洁面泡沫” - ingredient -“bisabolol红没药醇”). 

There are three types of newly added links: \textit{cause}, which relates scenario to problem (e.g., ``changeover period between autumn and winter 秋冬换季'' - cause - ``dry skin 皮肤干''); \textit{need}, which relates problem to POI (e.g., ``pimple 长痘痘'' - need - ``antiacne 清痘抑痘''); and \textit{satisfy}, which links property value and POI (e.g., ``bisabolol 红没药醇'' - satisfy - ``antiacne 清痘抑痘''). These links are established based on domain knowledge, while the dotted ones are added or predicted through KG completion. For instance, if a user has a problem ``pimple'', which needs ``antiacne'', we will add a preference link between the user and ``antiacne''. In addition, we use ``has\_image'' to link an image to a corresponding textual entity.

\subsection{KG Construction}

\begin{figure}[h]
	\centering
	\includegraphics[width=0.98\linewidth]{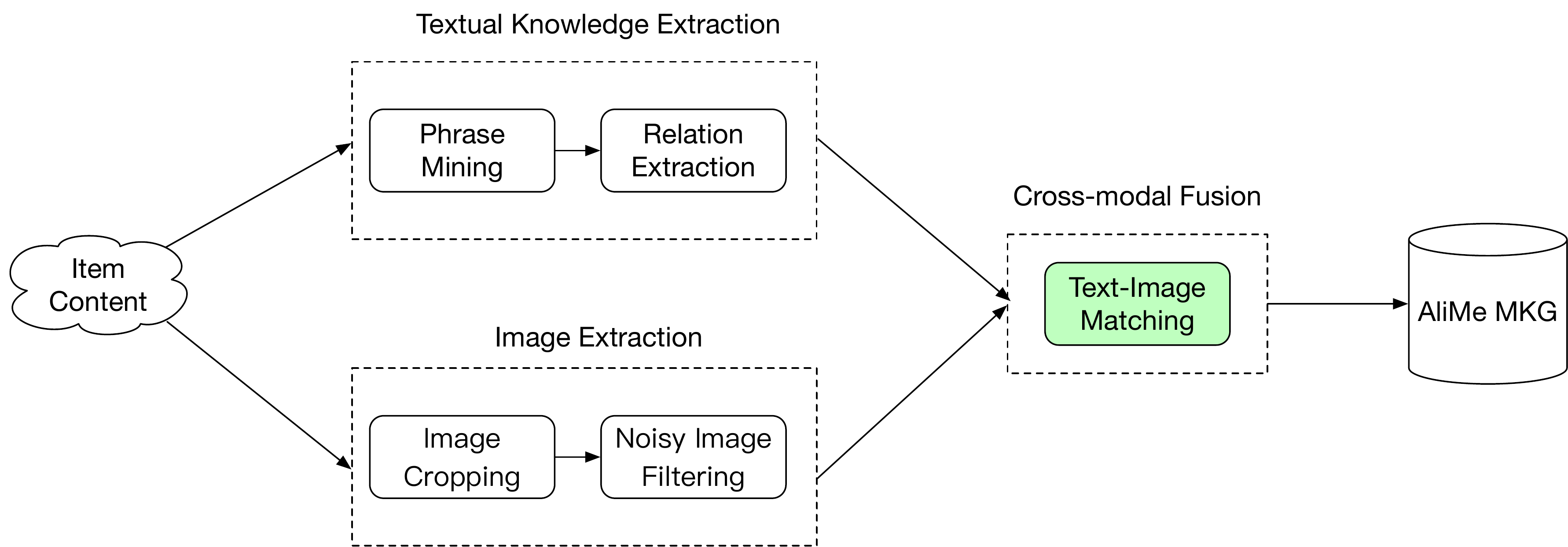}
	\caption{The construction process of AliMe MKG.}
	\label{fig:kgc}
\end{figure}

We present our KG construction process in Figure~\ref{fig:kgc}. In general, it includes three phases: textual knowledge extraction, image extraction and cross-modal fusion. 

\subsubsection{Textual Knowledge Extraction} 
As in AliMe KG~\cite{DBLP:journals/corr/abs-2009-11684}, $($Item, property, Property\_value$)$ triples are imported from the Alibaba product knowledge graph, POIs are extracted from E-commerce content (i.e., item detail pages and articles) using phrase mining and binary BERT classifier, and the relation between property values and POIs are acquired through relation extraction. Finally, the relation between an item and a POI can be inferred as follows: if an item has a property value that satisfy a specific POI, then the item will have that POI.

% Most textual triples can be collected from our previous KG. 
% For new items, we leverage named entity recognition and relation extraction to mining new triples to complete the MKG. Details of process of text knowledge extraction 
% can be found in previous work~\cite{DBLP:journals/corr/abs-2009-11684}, omitted due to the limited space.

\subsubsection{Image Extraction} 
We extract images mainly from item detail pages at present. Initially collected images can be overlong on size or convey time sensitive information (e.g.., a promotion will be offline after a certain time point). We first cut overlong images into pieces based on image content using edge detection, filter noisy images through image classification, optical character recognition (OCR) together with heuristic rules (e.g., the area proportion of OCR text, the number of OCR text blocks).

% The item detail pages are compose of some long images which express mixed information. 
% We have to process them and screen high quality images as candidates. First, we cut the long images into small images. 
% Next, we use OCR tool to recognize the OCR characters from the small images. 
% Then, we design heuristic rules to filter noisy images, for example, advertising or unattractive images. 
% The rules include the area ratio of OCR characters in the image, the length-to-width ratio.

\subsubsection{Cross-Modal Fusion} 
% We treat the linking of images to textual entities as a text-image matching problem. Inspired by the noticeable success of pre-trained language models (e.g., BERT~\cite{devlin2018bert}) in NLP, the pre-training for vision-language tasks has also attracted increasing attention in recent years. To overcome the limitation of pre-defined and mismatching categories of region detection, we choose to learn from pixels rather than bounding boxes to represent an image, as Pixel-BERT~\cite{DBLP:journals/corr/abs-2004-00849} does. We employ ResNet50~\cite{DBLP:conf/cvpr/HeZRS16}, which is freezed in our approach, as our backbone visual encoder, and initialize transformers using pre-trained StructBERT~\cite{DBLP:conf/iclr/0225BYWXBPS20} base model. Our model is similar to Pixel-BERT, but differs in that we use all extracted pixel features, instead of sampling, during our pre-training.

We treat the linking of images to textual entities as a text-image matching problem. Inspired by the noticeable success of pre-trained language models (e.g., BERT~\cite{devlin2018bert}) in NLP, the pre-training for vision-language tasks has also attracted increasing attention in recent years. On observing the limitation of pre-defined and mismatching categories of region detection~\cite{DBLP:journals/corr/abs-2004-00849}, and the effectiveness of directly applying Transformer to images~\cite{dosovitskiy2020image}, we choose to learn from image patches rather than bounding boxes, to represent an image. 

% our cross-model matching model is a two-stream Transformer network

\begin{figure}[h]
	\centering
	\includegraphics[width=0.96\linewidth]{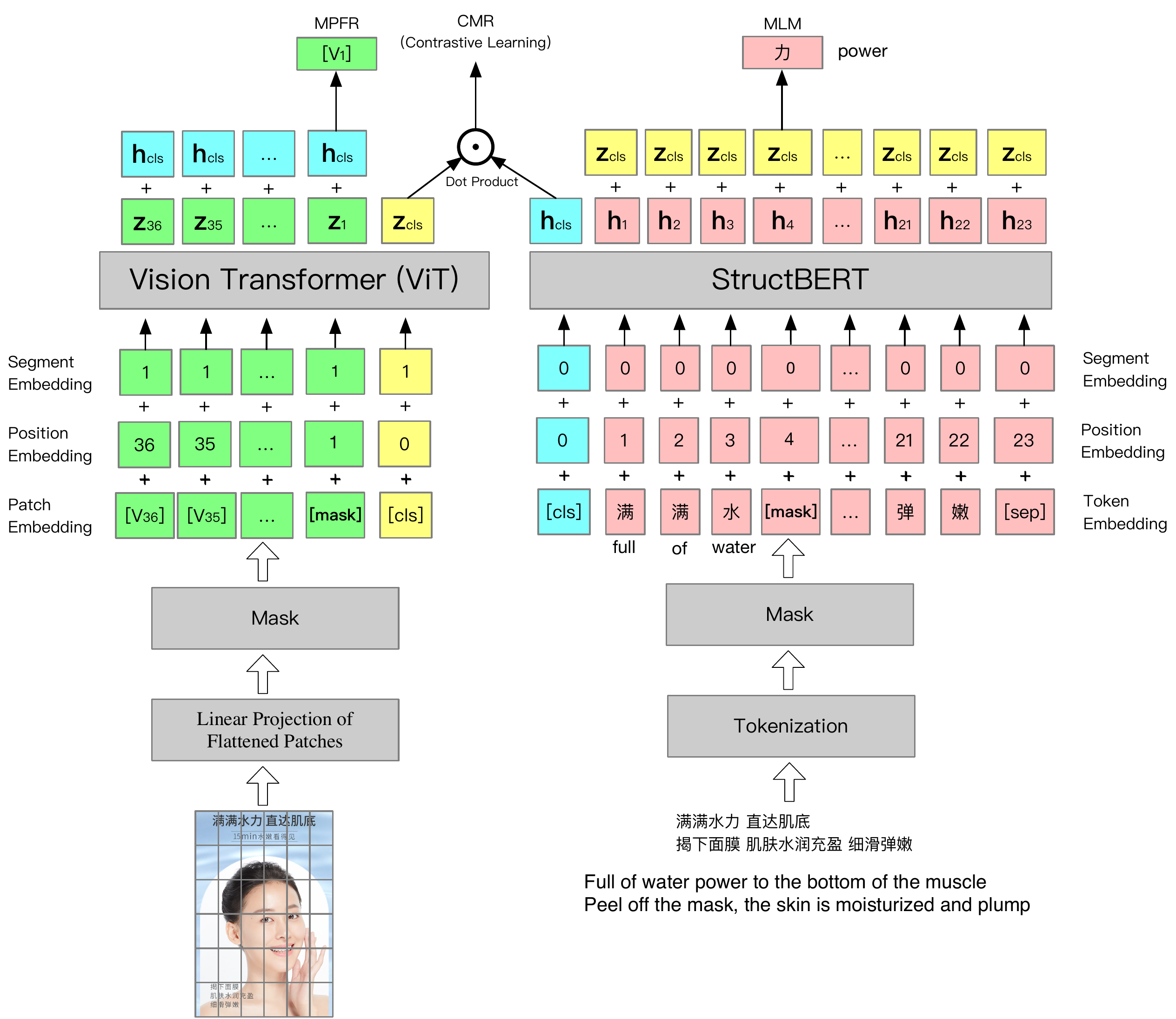}
	\caption{The overview of our cross-modal matching model.}
	\label{fig:mm_model}
	\vspace{-0.2cm}
\end{figure}

% \paragraph{Model Pre-training}
\textbf{Model Pre-training.}
As shown in Figure~\ref{fig:mm_model}, we follow LightningDOT~\cite{sun2021lightningdot} to adopt a two-stream Transformer architecture that removes the time-consuming cross-modal attention to accelerate inference. For image input (the left branch), we employ vision Transformer (ViT)~\cite{dosovitskiy2020image} as the feature extractor. Specifically, we first reshape an image $\bm{x} \in \mathbb{R}^{H \times W \times C}$ into a sequence of flattened 2-D patches $\bm{x}_{p} \in \mathbb{R} ^ {N \times (P^2 \times C)}$, where $(H, W)$ is the resolution of the original image, $C$ is the number of channels, $(P, P)$ is the resolution of each image patch and $N = HW/P^2$ is the number of patches, and then provide the sequence of linear embeddings $V_1, ... V_N$ of these patches along with corresponding segment and position embeddings as input to ViT. Moreover, to represent the whole image, we add a ``[CLS]'' token to the beginning of the patch list. For text input, we employ StructBERT~\cite{DBLP:conf/iclr/0225BYWXBPS20} as our backbone and follow the BERT convention to use the ``[CLS]'' token to represent the whole text.
%follow the BERT convention to sum the corresponding token, segment, and position embeddings as input representation.% 

%With feature extracted, we use light-weight dot-production for final matching. To accelerate inference, we adopt the two-stream model architecture as in LightningDOT~\cite{sun2021lightningdot} through removing the time-consuming cross-modal attention, and using lightweight dot-production for matching
% ~\cite{sun2021lightningdot}

We pre-train our model with a large amount of image-sentence pairs and three objectives, namely Masked Language Modeling (MLM), Masked Patch Feature Regression (MPFR) and Cross-modal Retrieval (CMR). In the MLM task, we consider the paired image $i$ as complementary information when reconstructing masked tokens in a sentence $t$ through adding the embedding of the whole image to that of the masked tokens. In the MPFR task, our model learns to regress the output of each masked patch to its visual feature through an MSE (mean square error) loss, and exploits the global text representation when reconstructing masked patches in a similar spirit. Finally, in the CMR task, we use dot product to measure the similarity score between a text $t$ and an image $i$. Specifically, we calculate the dot product between the embedding of ``[CLS]'' tokens respectively from ViT and StructBERT. To better capture the supervision signal, we employ a bi-directional variant of contrastive loss that is proposed in LightningDOT~\cite{sun2021lightningdot}: give a batch of $m$ matched image-text pairs $(i, t)$, we treat image $i$ as the query and the other $m-1$ texts $\{t_2, t_3, . . . , t_m\}$ as negatives, and perform similarly by taking text $t$ as the query.

To perform pre-training, we collect more than 6 million image-sentence pairs from items' detail pages. Operationally, the pairs are constructed through pairing an image with its OCR text. Heuristic rules are also employed to prune low quality text and images.

% In general, our pre-training includes two stages: textual pre-training with domain corpus and vision-language pre-training with text-image pairs. For textual pre-training, we collect 14 million chat logs from our customer service chatbot~\cite{li2017alime} in the E-commerce domain and use them to further pre-train StuctBERT with the Masked Language Modeling (MLM) task. For cross-modal pre-training, we collect 6 million image-sentence pairs from items' detail pages, and utilize them to train the domain-specific StructBERT with three pre-training tasks: MLM, Image-Text Matching (ITM) and Masked Region Feature Regression (MRFR). 

% Specifically, image-sentence pairs are constructed through pairing an image and its OCR text. Heuristic rules are also employed to prune low quality text or images. For negatives, we randomly select two items $I_a, I_b$ that are in the same category, and pair an image of $I_a$ with the paired text of another image that belongs to item $I_b$. For pre-trainng tasks, the first two are adopted from Pixel-BERT, and the last one is from UNITER~\cite{DBLP:conf/eccv/ChenLYK0G0020}. In the MRFR task, our model learns to regress the output of each masked pixel to its visual features through an MSE (mean square error) loss.

% \paragraph{Model Fine-tuning}
\textbf{Model Fine-tuning.}
For the downstream text-image matching task, we fine-tune the pre-trained multi-modal model with task-specific datasets. Finally, we will obtain triples that describe the ``has\_image'' relation between textual and image entities through model matching.

% For the downstream text-image matching task, we fine-tune the pre-trained multi-modal model with task-specific datasets. During fine-tuning, the input sequence is formulated as ``[CLS] $text_{to\_match}$ [SEP] $text_{ocr}$ [SEP] $image$'', where ``$text_{ocr}$'' is the text extracted from the image itself, and will be empty if there is no OCR text. Finally, we will obtain triples that describe the ``has\_image'' relation between textual and image entities through model matching.

\subsection{The Performance of Cross-Modal Matching}
We compared our cross-modal matching (aka CMM) approach with Pixel-BERT~\cite{DBLP:journals/corr/abs-2004-00849}, which employs ResNet50~\cite{DBLP:conf/cvpr/HeZRS16} to extract image features, and feeds the text and visual features into a single-stream Transformer network for image-text matching. For fair comparison, we also pre-trained Pixel-BERT with the same amount of image-text pairs in E-commerce. Instead of sampling pixels, we use all extracted pixel features during our pre-training. Also, we compared the linear projection of patches with ResNet50 on image feature extraction when using ViT as the image encoder in our approach.

\begin{table}[h]
  \vspace{-0.1cm}
  \caption{Experimental results of cross-modal matching}
  \vspace{-0.1cm}
  \label{tab:exp_mm_matching}
  \begin{adjustbox}{max width=0.9\linewidth}
  \begin{tabular}{llll}
    \toprule
    Model& AUC & Inference & Image Feature Extraction\\
    \midrule
    Pixel-BERT & 0.9853 & 1$\times$ & 1$\times$ \\
    CMM\_ResNet & 0.9846 & 55.18$\times$ & 0.81$\times$ \\
    \textbf{CMM\_Patch} & \textbf{0.9861} & \textbf{55.18$\times$} & \textbf{1.61$\times$} \\
  \bottomrule
\end{tabular}
\end{adjustbox}
\end{table}

We evaluated the three methods on an image-text matching dataset in the Beauty domain, where each text has 10 candidate images. We extract image features offline in advance, and then perform image-text matching for each coming text query. We run each experiment 3 times and show the average $AUC$, inference speed and feature extraction speed in Table~\ref{tab:exp_mm_matching}. We can see that our approach not only achieves the best performance (AUC=0.9861) but also brings substantial speed improvement (55.18$\times$ for online inference time and 1.61$\times$ for offline image feature extraction).

\subsection{KG Statistics}
Our multi-modal KG is ongoing. Currently, our MKG covers three vertical domains, namely Clothing, Beauty and Snacks. It has accumulated 400$^+$ scenarios, 1K$^+$ user problems, 500K$^+$ POIs, 2K$^+$ ``Scenario - cause - Problem'' triples, 12K$^+$ ``Problem - need - POI'' triples, 500K$^+$ ``Property\_value - satisfy - POI'' triples, 300K$^+$ items, and 28M$^+$ associated images. The triples except items and images have been completely checked by crowd-sourcing, hence their quality can be ensured. The spot check on cross-modal matching shows that the accuracy is of high quality and can be applied in practice.

%1$M+$ ``Property\_value - satisfy - POI'' triples, and 28$M^+$ images that are associated with property values and POIs

% has accumulated 300K+ items, 1M+ “Property_value - satisfy- POI” triples, and 28M+ images that are associated with property values and POIs.

\subsection{KG Example}
% We show an excerpt of our multi-modal knowledge graph in Figure~\ref{fig:kge}. As it shows, the scenario ``Winter'' causes the problem ``Dry skin'', which needs ``relieve skin''. A ``facial mask'' item contains as its ingredient ``Centella'', which is able to help to ``relieve skin'', and hence is fit for corresponding users.

\begin{figure}[h]
	\centering
	\includegraphics[width=0.96\linewidth]{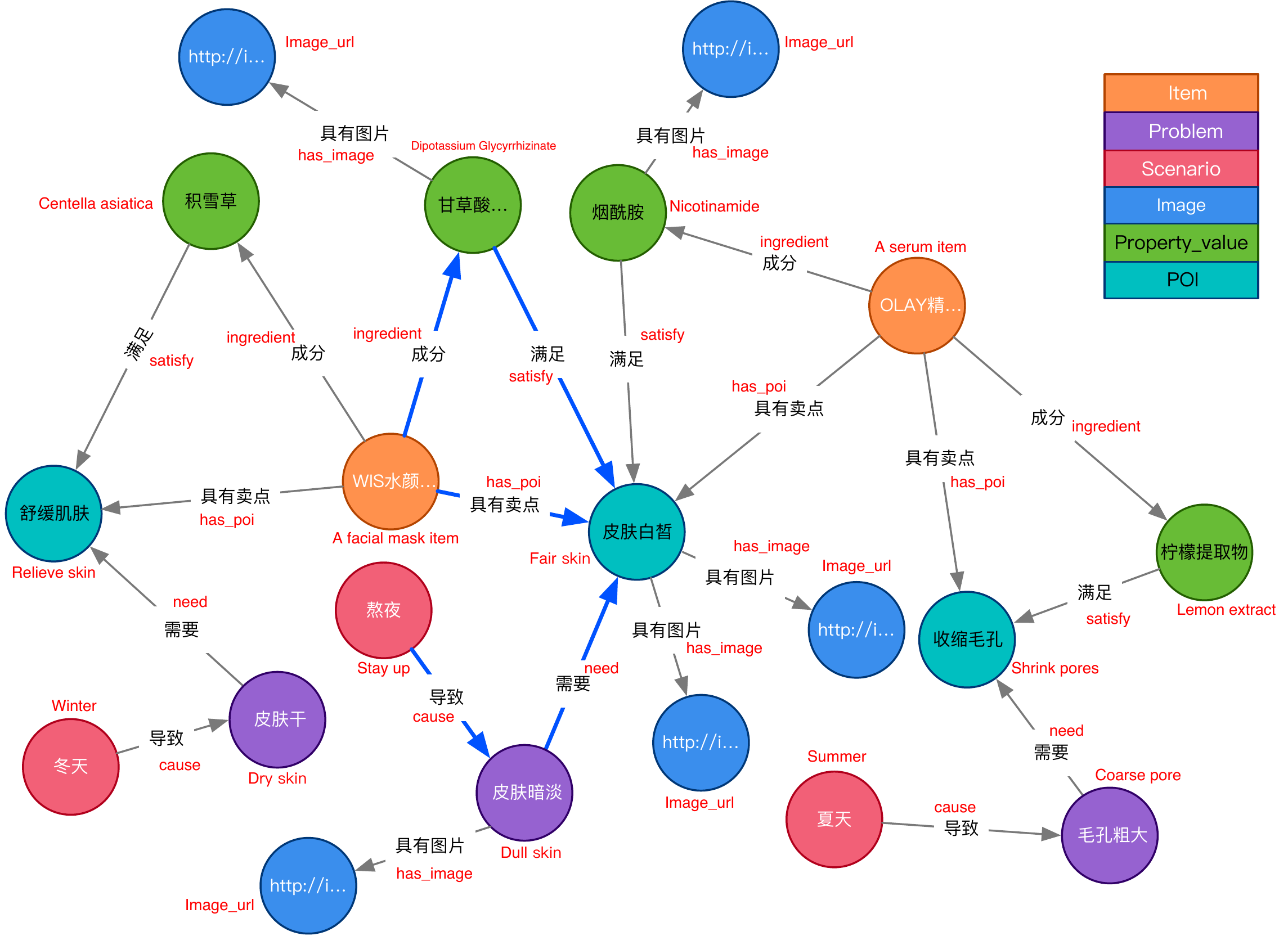}
	\caption{An excerpt of our multi-modal knowledge graph.}
	\label{fig:kge}
	\vspace{-0.5cm}
\end{figure}

We show an excerpt of our multi-modal knowledge graph in Figure~\ref{fig:kge}. As it shows, the scenario ``Stay up late 熬夜'' causes the problem ``Dull skin 皮肤暗沉'', which requires ``Fair skin 皮肤白皙''. A ``facial mask 面膜'' item contains as its ingredient ``Dipotassium glycyrrhizinate 甘草酸二钾'', which is able to help to ``Fair skin'', and hence is fit for corresponding users.

\section{KG Application: Live Assistant}
\begin{figure}[h]
	\centering
	\includegraphics[width=0.9\linewidth]{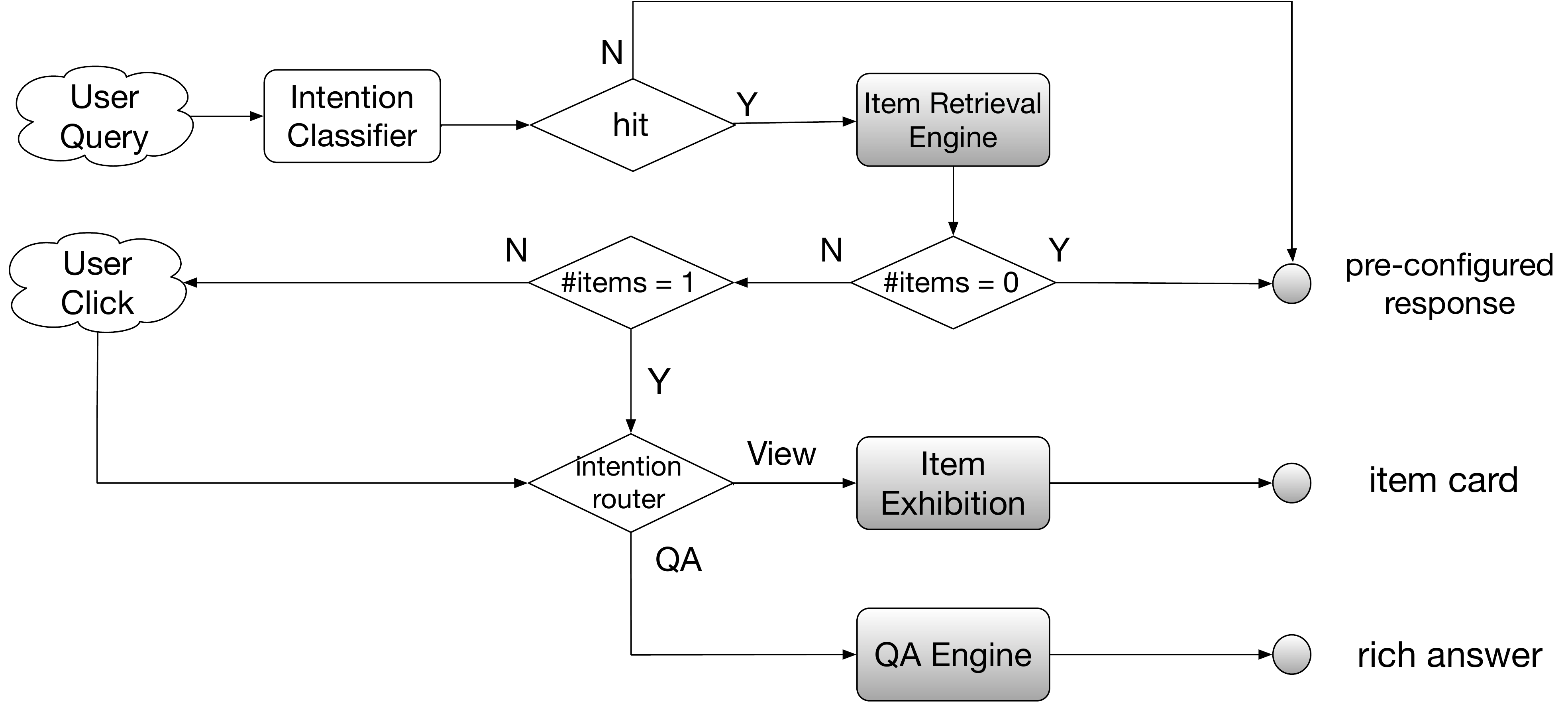}
	\caption{The overall processing flow of Live Assistant.}
	\label{fig:processing}
\end{figure}

We have built an online live assistant based on our MKG for users to skim over item list, view item details and ask item-related questions. We show the overall processing flow of customer inputs in Figure~\ref{fig:processing}. Given a user query $q$, our online system first conduct intention identification. If a user is requesting for viewing item, e.g., ``Can I see the lipstick?", s/he will be directed to our \textit{Item Exhibition} component that displays item cards. If the user is asking item-related question, e.g., ``What is the size of the T-shirt?'', s/he will be replied by our \textit{QA Engine} module. If the query is irrelevant to products, our system will respond with a pre-configured answer. 

\paragraph{Item Retrieval Engine.} 
Given a customer query $q$ that is related to product(s), the engine will parse the query, and search for and rank related items. If there are more than one item, it will return the ranked list, and ask users for their selection. To enable our model to better understand products, we employ a offline NER model to identify the key aspects of items from their titles and detail pages, e.g., category, brand, functionality, etc., and add the identified semantic types into product profiles in our KG. During item search, we employ FLAT~\cite{DBLP:conf/acl/LiYQH20}, a light-weight NER model to identify pre-defined semantic types from the query $q$, and calculate an enhanced similarity score between $q$ and each item based on their tokens and identified semantic types. 

\paragraph{Item Exhibition.} 
Once a specific item is chosen, our live assistant will retrieve a rich set of information about the item from our multi-modal KG, and properly display them in a pop-up window. Roughly, item information is organized into three categories: appearance, POI and comment. The core elements of our MKG, including scenarios, POIs, item images and property values, are exhibited according to the pre-defined structure.

%All of the core elements of our MKG, including property values, POIs and item images, are well supported at the front end. 

% As we are working on encouraging customers to purchase within broadcasting rooms, then the richness and attractiveness of information is of key importance. That is, our MKG is positioned as providing cognitive product profiles, through which customers are able to truly understand a product: why they need, when and where can they use, how to use, and what is the effect. 

% As shown in Figure\ref{fig:app} (b), the textual POI entities are shown as short tags to describe key features of the item, 
% the hot textual Property relations are provided for customers to further click, the image entities are arranged for customers to slide and view.

% Once the specific product is confirmed, this component will show the organized rich information about the item obtained from MKG.
% As illustrated in Figure\ref{fig:app} (b), the textual POI entities are shown as short tags to describe key features of the item, 
% the hot textual Property relations are provided for customers to further click, the image entities are arranged for customers to slide and view.

\paragraph{QA Engine.} 

To help an anchor to answer questions, we designed a hybrid approach that employs both KBQA (question answering over knowledge base) and DeepQA (question answering over frequently asked question).
In KBQA, we adopt an NER model to identify properties from customer questions, and retrieve corresponding property values and associated images from our MKG. To improve readability, we synthesize a natural language sentence instead of delivering a single property value. If there is no identified property in $q$ or no answer in the MKG, we use a text matching model to find the most similar FAQ and return the corresponding answer.

% To help an anchor to answer questions, we designed a hybrid approach that employs both KBQA (question answering over knowledge base) and DeepQA (question answering over frequently asked question through deep learning text classification and text matching).
% In our scenario, KBQA is preferred than DeepQA due to: 1) KBQA is able to perform fine-grained semantic parsing, based on which (e.g., specific properties like size chart, shape) we are able to display corresponding item information; 2) the majority of knowledge in structured, and the content in our MKG is more plentiful than that in FAQs.

% Specifically, we adopt an NER model to identify properties from customer questions, 
% and retrieve corresponding property values and associated images from our MKG. To improve readability, we synthesize a natural language sentence instead of delivering a single property value. We currently generate sentences based on textual property value through templates, and are going to explore generative models with pre-training. If there is no identified property in $q$ or no answer in the MKG, we use a text matching model to find the most similar FAQ and return the corresponding answer.

% \subsection{KG Application: Short Video Production}
% We have also explored to utilize our MKG for short video production. 

\section{Demonstration}

Our system has been launched online in the Taobao app, and currently serves hundreds of thousands of customers per day~
% \footnote{Our video demonstration is available at: \url{https://youtu.be/UZgRCUrNvaU}}.

\begin{figure}[t]
	\centering
	\includegraphics[width=0.96\linewidth]{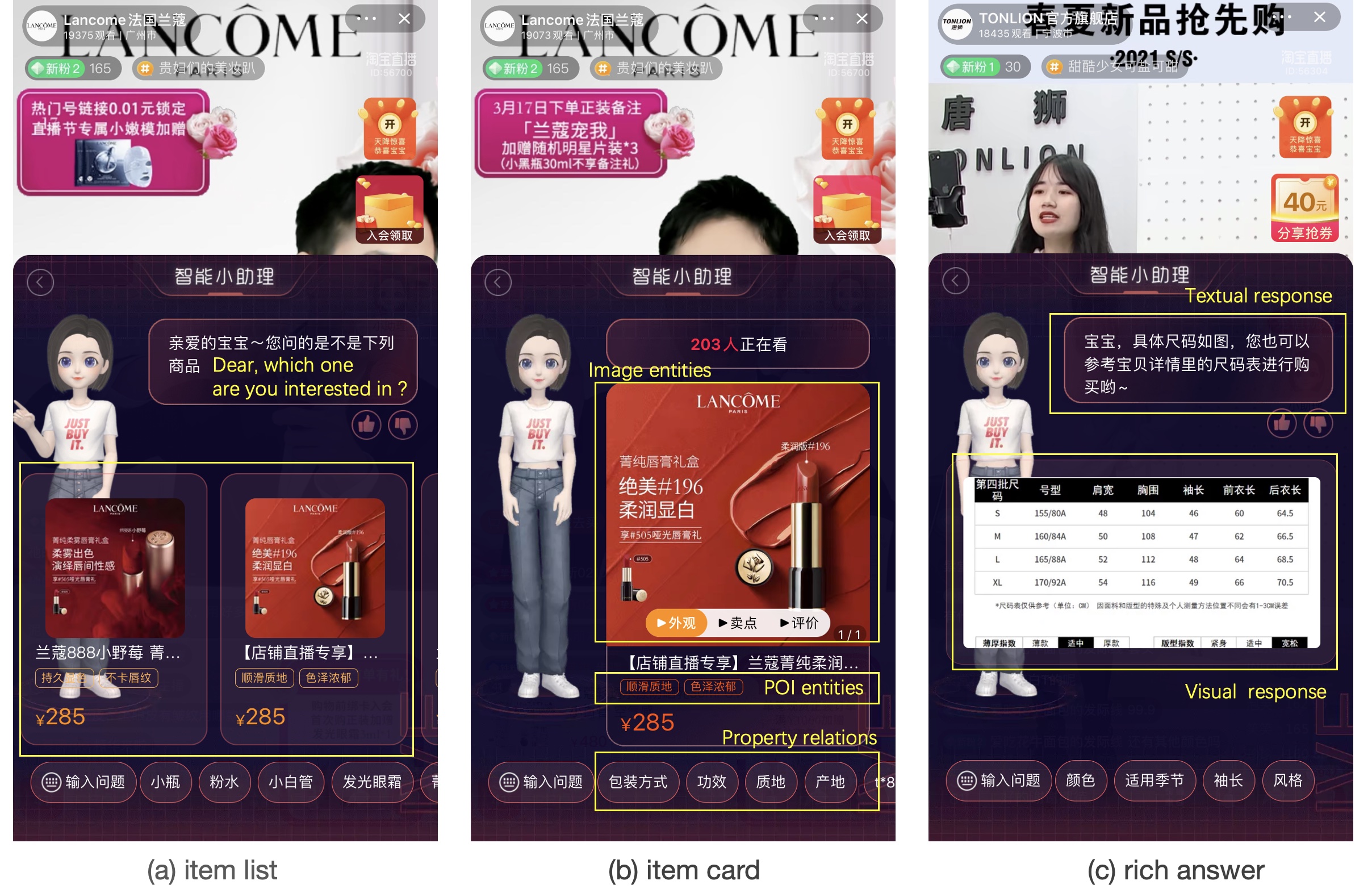}
	\caption{A demonstration of Live Assistant.}
	\label{fig:app}
	\vspace{-0.2cm}
\end{figure}

We demonstrate the key features of our system in Figure~\ref{fig:app}. Figure~\ref{fig:app} (a) shows a list of items that are retrieved based on a user query ``Can I see the lipstick?'' and shown in a pop-up window. Users are able to skim over the item list and choose the one that interests them most. Once an item is chosen, the important images, POIs and properties of the item will be exhibited, as shown in figure~\ref{fig:app} (b). Figure~\ref{fig:app} (c) demonstrates the richness of our KBQA answer, which contains a textual description and a size chart, and answers a question about the size of a specific T-shirt.

% We demonstrate the key features of our system in Figure~\ref{fig:app}. Figure~\ref{fig:app} (a) shows a list of items that are retrieved based on a user query ``Can I see the lipstick?''~\footnote{The query is currently not present.} and shown in a pop-up window. Users are able to skim over the item list and choose the one that interests them most. Once an item is chosen, the important images, POIs and properties of the item will be exhibited, as shown in figure~\ref{fig:app} (b). Figure~\ref{fig:app} (c) demonstrates the richness of our KBQA answer, which contains a textual description and a size chart, and answers a question about the size of a specific T-shirt.

% Note that the three kinds of information are independent, and there is no correspondence between them. 

% \begin{figure}[t]
% 	\centering
% 	\includegraphics[width=\linewidth,height=0.888\linewidth]{files/multi_picture.png}
% 	\caption{A demonstration of short video.}
% 	\label{fig:multi_picture}
% \end{figure}

% \subsection{KG Example}

% We show an excerpt of our multi-modal knowledge graph in Figure~\ref{fig:svdemo}. As it shows, the scenario ``Winter'' causes the problem ``Dry skin'', which needs ``relieve skin''. A ``facial mask'' item contains as its ingredient ``Centella'', which is able to help to ``relieve skin'', and hence is fit for corresponding users.

\begin{figure}[h]
	\centering
	\includegraphics[width=0.96\linewidth]{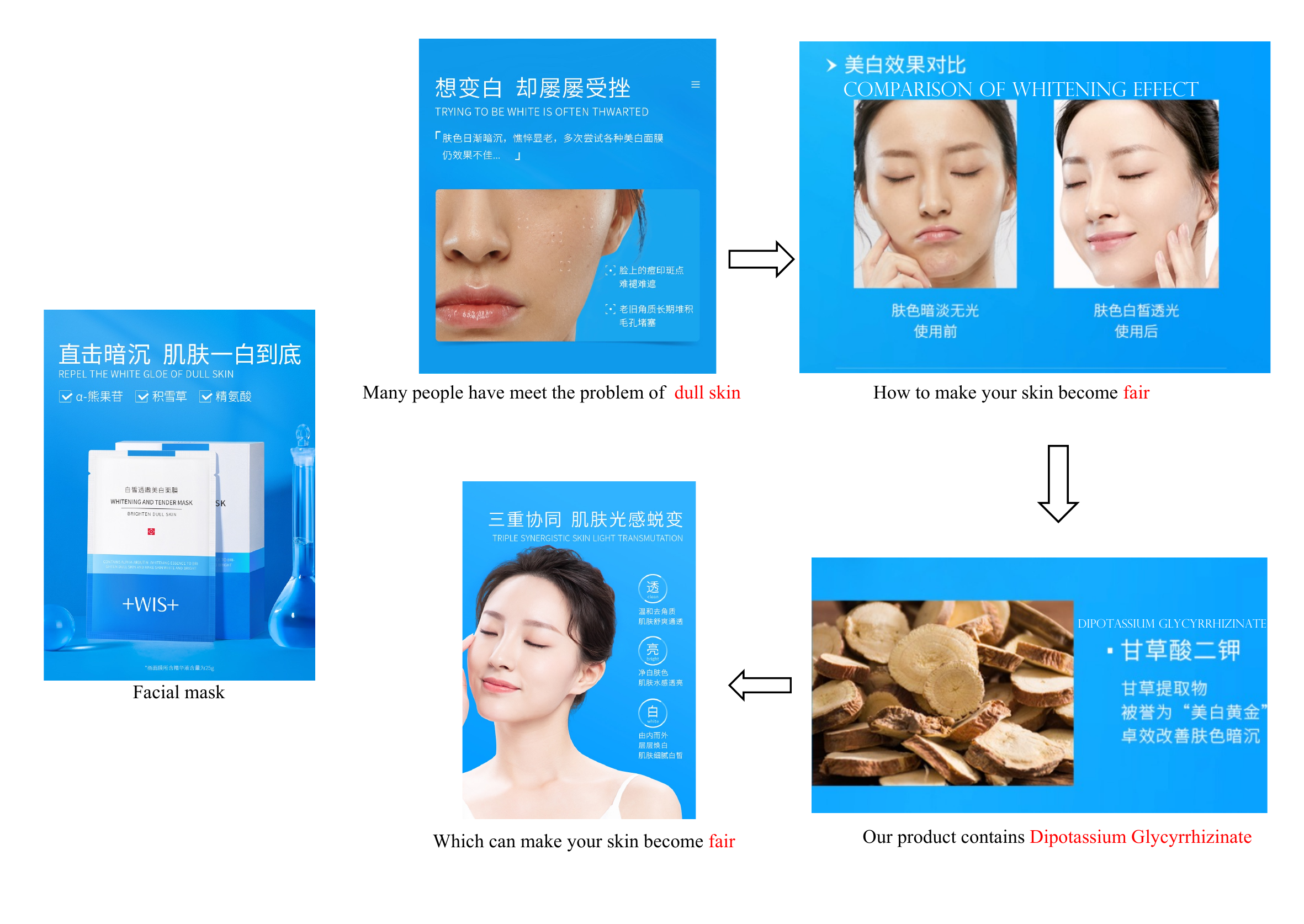}
	\vspace{-0.3cm}
	\caption{A demonstration of short video generation based on our MKG.}
	\label{fig:svdemo}
	\vspace{-0.2cm}
\end{figure}

We also show an innovative application of our multi-modal KG in short video production in Figure~\ref{fig:svdemo}. By following the cognitive path in Figure~\ref{fig:kge}, we generate a nature language utterance for each node, and organize the associated images according to the given order, and finally produce a short video through using templates. The generated short videos can be played in the item card as in Figure~\ref{fig:app} (b). Such knowledge-based short videos tell the core selling points of a product item in an attractive and cognitive manner, hence are more convincing on affecting customers' buying decision. Moreover, with our knowledge graph, the productivity of such knowledge-based video producing can be largely improved, making its large-scale application feasible.

\section{Conclusion}
In this work, we present AliMe MKG, a multi-modal knowledge graph that aggregates rich product information, and introduce its innovative applications in live-streaming E-commerce. Providing cognitive product profiles to customers is of value and challenging. Many interesting problems, such as acquiring encyclopedic or scenarized knowledge, personalized exhibition, knowledge enhanced text generation will be further explored to enrich our multi-modal KG and polish our online live assistant.

%In the future, we plan to further explore concept-level cross-modal matching that may need external knowledge, and knowledgeable text generation, to enrich our multi-modal KG and polish our online live assistant. 

% As we are working on encouraging customers to purchase within broadcasting rooms, then the richness and attractiveness of information is of key importance. That is, our MKG is positioned as providing cognitive product profiles, through which customers are able to truly understand a product: why they need, when and where can they use, how to use, and what is the effect.

%Many interesting problems remain open.
% We introduce the construction process and describe the cross-modal fusion in detail. 
% Based on Alime KG, we propose an online live assistant launched in the Taobao app, giving product exhibition and question answering for customers. 

% In this work, we present Alime KG, a multimodal knowledge graph that aggregates rich information about items for live streaming in E-commerce. 
% We introduce the construction process and describe the cross-modal fusion in detail. 
% Based on Alime KG, we propose an online live assistant launched in the Taobao app, giving product exhibition and question answering for customers. 


\bibliographystyle{ACM-Reference-Format}
\bibliography{cikm2021}

\end{CJK*}
\end{document}